\documentclass[prb,aps,twocolumn,showpacs,superscriptaddress,10pt]{revtex4-1}
\usepackage[utf8]{inputenc}
\usepackage{amssymb,amsmath}
\usepackage{graphicx}
\usepackage[T1]{fontenc}
\usepackage{multirow}
\usepackage{color}
\usepackage[english]{babel}
\usepackage{hyperref}

\newcommand{\ket}[1]{ \left| #1 \right \rangle}

\begin{document}
\title{\textbf{Anisotropic spin-valley coupling in SiMOS and Si/SiGe quantum dots}}
\author{N. Tobias Jacobson}
\thanks{These authors contributed equally to this work.}
\email[Corresponding author: ]{ntjacob@sandia.gov}
\affiliation{Sandia National Laboratories, Albuquerque, NM 87185, USA}
\author{Natalie D. Foster}
\thanks{These authors contributed equally to this work.}
\affiliation{Sandia National Laboratories, Albuquerque, NM 87185, USA}
\author{Ryan M. Jock}
\thanks{These authors contributed equally to this work.}
\affiliation{Sandia National Laboratories, Albuquerque, NM 87185, USA}
\author{Martin Rudolph}
\affiliation{Sandia National Laboratories, Albuquerque, NM 87185, USA}
\author{Andrew M. Mounce}
\affiliation{Sandia National Laboratories, Albuquerque, NM 87185, USA}
\author{Daniel R. Ward}
\affiliation{Sandia National Laboratories, Albuquerque, NM 87185, USA}
\author{Malcolm S. Carroll}
\affiliation{Sandia National Laboratories, Albuquerque, NM 87185, USA}
\author{Dwight R. Luhman}
\affiliation{Sandia National Laboratories, Albuquerque, NM 87185, USA}

\begin{abstract}
While bulk silicon has long been understood to exhibit relatively weak spin-orbit coupling (SOC), confinement of electrons to quantum dots (QDs) at a silicon heterointerface results in significantly larger SOC. This is a concern for electron spin qubit performance, as intravalley and intervalley SOC can significantly perturb the operation of electron spin qubits. While these interactions can be harnessed to drive coherent rotations in a singlet-triplet qubit, coupling to low-lying excited valley states can lead to undesirable spin relaxation when valley splitting is on resonance with the Zeeman energy. In this work, we measure the angular dependence of the interfacial spin-orbit interaction as a function of the direction and magnitude of an applied external magnetic field in SiMOS and Si/SiGe heterostructures, two common material platforms for silicon spin qubits. We construct a physical model that accurately infers intra- and inter-valley SOC physics from fits to the data, allowing for a direct comparison between these two material systems. For the devices measured we find that, while the $g$-factor differences are comparable, the SiMOS QDs exhibit an order of magnitude larger spin-valley coupling than for Si/SiGe. Moreover, we find that the angular dependence of the spin-valley coupling is similar for both devices, with similar magnetic field orientations minimizing the spin-valley coupling. Our work points towards operational schemes for optimizing spin-valley coupling to avoid or exploit this mechanism for qubit operation.
\end{abstract}

\maketitle

\label{sec:Introduction}
Silicon's promise as an excellent material for hosting spin qubits stems from ready isotopic purification---the ability to nearly eliminate spinful background nuclei---and the relatively weak spin-orbit coupling (SOC) experienced by conduction band electrons in bulk silicon as compared with other semiconducting materials such as GaAs\cite{Winkler2003}. However, experiments in recent years have indicated that confining electrons to quantum dots (QDs) at the Si/SiO$_2$ interface \cite{Veldhorst2015,Jock2018,Harvey-Collard2019,Ferdous2018,Fogarty2018,Tanttu2019} or within Si wells in between SiGe barriers \cite{Cai2023,Foster2025} generates SOC strong enough to affect qubit operation. The broken crystal symmetry at a silicon interface allows for spin-orbit physics such as Dresselhaus SOC that is forbidden by symmetry in the bulk, contributing to significant variation of electronic $g$-factor from one QD to the next. This $g$-factor difference between neighboring dots is sufficient to drive singlet-triplet (ST) rotations \cite{Jock2018,Harvey-Collard2019,Jock2022} and, in single-spin implementations using electron spin resonance, facilitates single-qubit addressability \cite{Veldhorst2015aa,Hwang2017,Ferdous2018,Ferdous2018b}. On the other hand, SOC-induced $g$-factor differences may generate errors for qubit encodings such as exchange-only qubits that depend on magnetic field uniformity~\cite{Burkard2023}.

Furthermore, SOC between valley eigenstates with opposite spin orientations leads to enhanced electron spin relaxation (shorter spin T$_1$) when the electronic Zeeman splitting is near-resonant with the valley splitting \cite{Yang2013,Hao2014,Huang2014,Hwang2017,Corna2018,Hollmann2020}. Provided valley splittings were consistently large, operating with moderate magnetic fields would mitigate this error mechanism. However, significant variability of the valley splitting has been observed across measured devices \cite{PaqueletWuetz2022,Volmer2024,Marcks2025}. This variability may complicate other qubit operations, such as spin shuttling \cite{Langrock2023}, due to exposure to spin-valley-mediated fast relaxation channels or unintentional transitions into excited valley states. This need for consistently large valley splitting motivated theoretical \cite{Zhang2013,Yi2022,Losert2023,Lima2024} and experimental \cite{McJunkin2022,Stehouwer2025} investigations into alternative device growth or control parameters to decrease the likelihood of encountering low-energy valley states. Despite this risk to relaxation time from spin-valley coupling, previous demonstrations have shown that this intervalley SOC mechanism can itself be leveraged to provide a drive mechanism for coherent evolution \cite{Harvey-Collard2019,Liu2021,Jock2022,Cai2023}.

In this work, we investigate the interplay between intrinsic intravalley (i.e. $g$-factor differences) and intervalley (i.e. spin-valley coupling) spin-orbit physics in silicon quantum dots in two spin qubit material systems, SiMOS and Si/SiGe. To do so, we estimate the effective magnetic field gradient between QDs through measurements of the free induction decay of a singlet-triplet (ST) qubit in the S/T$_0$ encoding as a function of magnetic field orientation and strength. We find a complex field dependence of ST rotation frequency on applied magnetic field magnitude and orientation that we explain with a simple model. This model relates the functional forms of observed magnetic field-dependent evolution frequencies to three physical scenarios of $g$-factor variation and spin-valley coupling. We find that the spin-valley coupling in the SiMOS device is roughly an order of magnitude larger than for the Si/SiGe device. At the same time, the valley splittings in the SiMOS device are significantly larger than for the Si/SiGe device. The magnetic field orientation dependencies that we find for both $g$-factor difference and spin-valley coupling point to strategies that may be used to enhance or mitigate these effects on qubit operation.

Our manuscript is organized as follows: In Sec. \ref{sec:Results} we describe the devices and our measurement protocol, in Sec. \ref{sec:Discussion} we present our model and discuss our results, and we conclude with Sec. \ref{sec:Conclusion}. We provide further details of the model and discuss additional measurements in the Appendix.

\section{Results}
\label{sec:Results}

\begin{figure}[!h]
	\centering
	\includegraphics[width=0.45\textwidth]{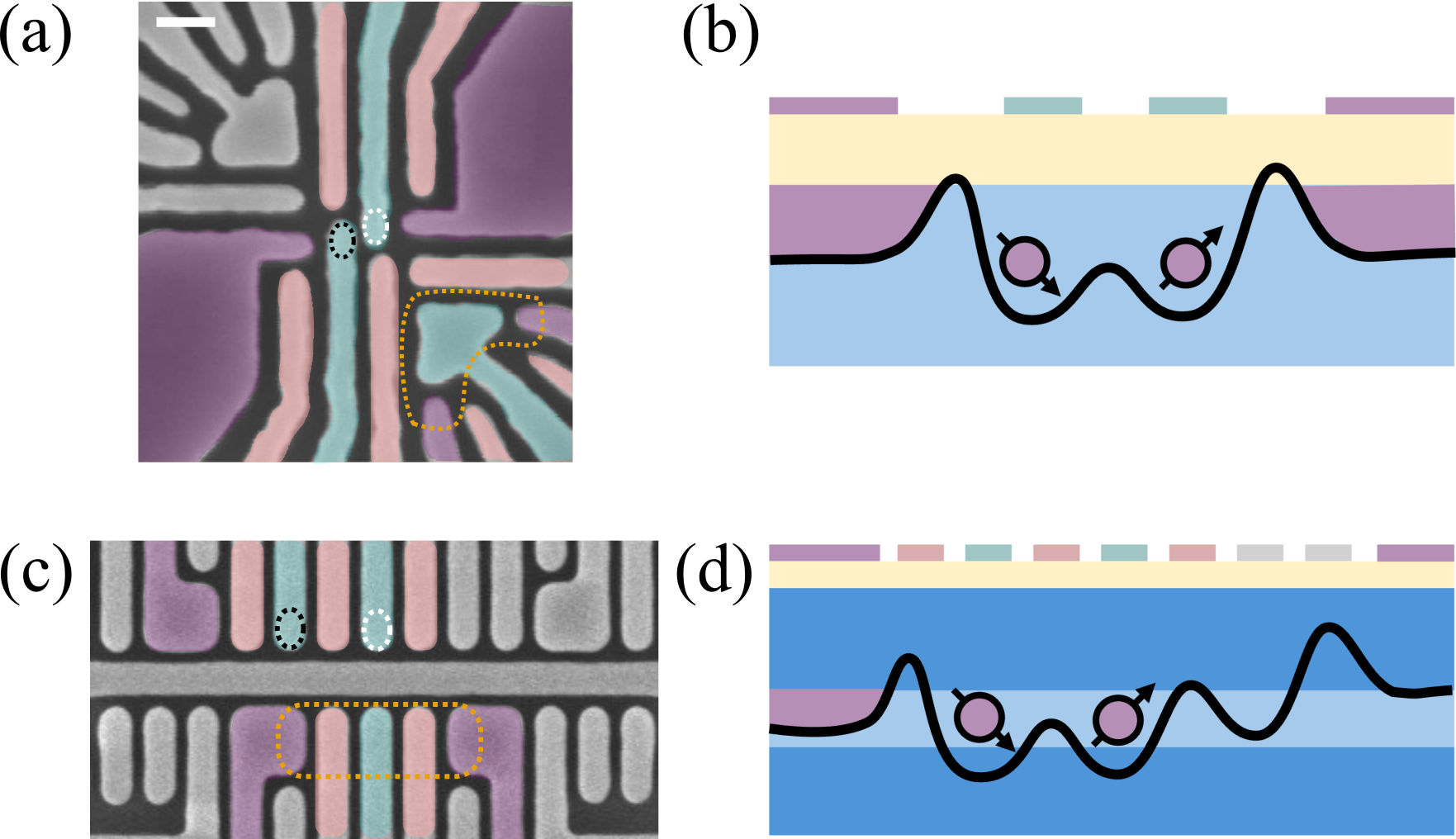}
	\caption{\textbf{Schematics of measured devices.} False-color scanning electron microscopy (SEM) images of \textbf{(a)} SiMOS and \textbf{(c)} Si/SiGe devices. The crystallographic axis [110] ([1$\bar{1}$0]) is oriented along the inter-dot $x$-axis of the SEM images for the SiMOS (Si/SiGe) device, respectively. Relevant gates shown are accumulation gates (purple) underneath which a two-dimensional electron gas forms, barrier gates (pink), and plunger gates (blue). Quantum dots Q$_1$ (black dashed) and Q$_2$ (white dashed) are formed underneath plunger gates. Proximal charge sensors infer the electron spin states (orange dashed). Gates not immediately relevant to this work are shown in light gray. The white scale bar represents 100 nm and only applies to the SiMOS device. Side-cut sketches in the region where the qubit is formed in \textbf{(b)} SiMOS and \textbf{(d)} Si/SiGe devices. In SiMOS, electrons are vertically confined to the interface of enriched Si (light blue) and SiO$_2$ (light yellow). In Si/SiGe, electrons are vertically confined within the enriched Si quantum well (light blue) in between two natural SiGe barriers (dark blue). The barrier is isolated from the metal gates by a dielectric layer (yellow).
    }
	\label{fig:DeviceIntro}
    
\end{figure}

The two devices we measure in this work are based on distinct silicon QD material stacks, SiMOS and Si/SiGe, shown in Fig. \ref{fig:DeviceIntro}. The double quantum dot SiMOS device was fabricated at Sandia National Laboratories (reported on in Ref. [\onlinecite{Jock2022}]) and consists of a thermal oxide layer of SiO$_2$ interfaced with isotopically enriched silicon (500 ppm $^{29}$Si) in which two QDs are formed. The triple quantum dot Si/SiGe device was industrially fabricated at Intel Corp.'s quantum device foundry (similar to Ref [\onlinecite{Foster2025}]) and consists of an approximately 5 nm thick isotopically enriched silicon (800 ppm $^{29}$Si) quantum well sandwiched between Si$_{0.7}$Ge$_{0.3}$ barriers having natural isotopic abundance. We discuss more details of the experimental setup in Appendix~\ref{sec:measurement}.

\begin{figure*}[t]
	\centering
	\includegraphics[width=0.75\linewidth]{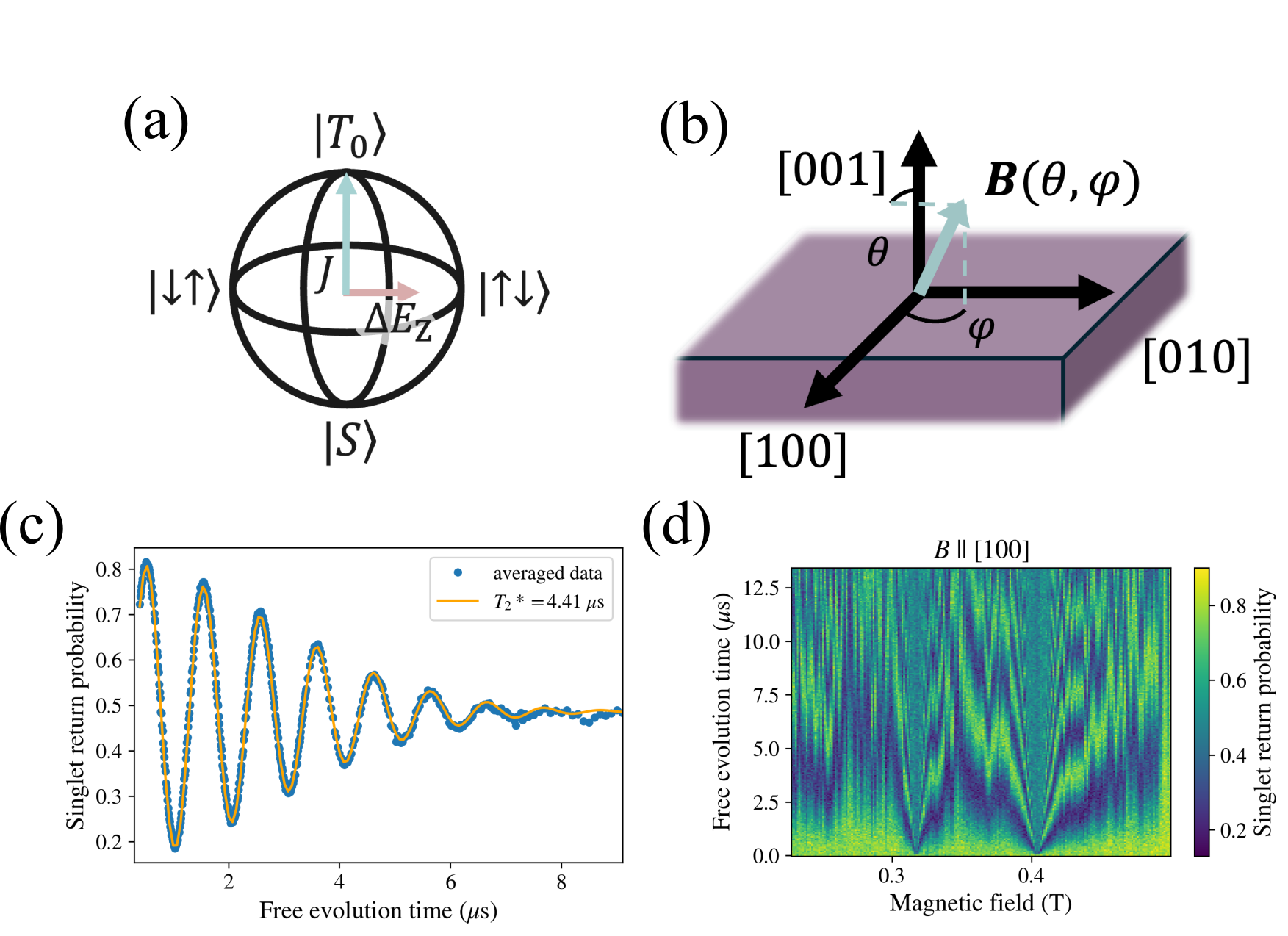}
	\caption{\textbf{Singlet-triplet free induction decay measurements.} \textbf{(a)} The ST Bloch sphere comprised of orthogonal control axes originating from the Zeeman energy difference due to SOC in pink ($\Delta E_\mathrm{Z}$) and exchange coupling due to wave function overlap in blue ($J$). The eigenstates of the magnetic field gradient are shown as $\vert \! \downarrow\uparrow\rangle$ and $\vert \! \uparrow\downarrow\rangle$ along the equatorial directions. \textbf{(b)} Schematic of the external magnetic field vector $\textbf{B} (\theta,\phi)$ relative to the crystallographic axes of Si. The elevation angle $\theta$ is defined relative to $[001]$, and $\phi$ is defined relative to $[100]$. \textbf{(c)} Example of free induction decay measurement on the Si/SiGe device, repeated and averaged over 30 minutes of data acquisition time with $\textbf{B}~\vert\vert~[110]=50$ mT. Data points in blue are fit to a Gaussian-decaying sinusoid in orange, resulting in an inhomogeneous dephasing time of $T_2^*=$4.41 $\mu$s. \textbf{(d)} Example raw data of a repeated free-induction decay measurement on the Si/SiGe device acquired while the magnetic field magnitude was ramped \textit{in-situ} along $[100]$. Each scan in variable evolution time was averaged over 100 shots. The corresponding FFT of these data is shown in Fig.~\ref{fig:ExperimentalMagnetospectroscopy_SiSiGe}a.}
	\label{fig:QubitOperation}
\end{figure*}

We use double QDs in each device to form a ST qubit, encoded into the singlet ($S$) and unpolarized triplet ($T_{0}$) states, with corresponding Bloch sphere shown in Fig.~\ref{fig:QubitOperation}a. We operate the device near the transition between (4,0) and (3,1) charge configurations, where $(n_1,~n_2)$ denotes the number of electrons in dots QD$_1$ and QD$_2$, respectively. Two electrons on QD$_1$ form a non-interacting spin-paired (filled-shell) singlet and the dynamics of the remaining two electrons can be electrically controlled by varying the detuning between the QDs. We first initialize the qubit by loading a $S$(4,0) ground state, then adiabatically separate one spin to form the $S$(3,1) spin state. For shallow detuning, there is significant electron wavefunction overlap between the top two electrons, one in each QD, and the exchange energy, $J$, is the dominant interaction. For two electrons that are sufficiently well-separated, $J$ is negligible and distinct Zeeman energies result from the differences in effective electron $g$-factor in each QD due to variations in their interfacial SOC \cite{Jock2018}, spin-valley coupling, and also from the local magnetic field gradient due to hyperfine coupling with the nuclear spin environment. These differences in Zeeman energy splitting between the two QDs, $\Delta E_\mathrm{Z}$, drive rotations between the $|S\rangle$ and $|T_0\rangle$ two-electron spin states. For measurement, we rapidly return the system to the (4,0) charge state, where Pauli spin blockade (PSB), combined with an enhanced latching mechanism \cite{Harvey-Collard2018a}, is used to read out the charge state and infer the spin state of the ST qubit. By varying the evolution time in the separated S(3,1) configuration, we measure the ST rotation frequency. In our modeling of this system, we make the filled-shell approximation and treat the two QDs as a system of two electron spins.

To capture the angular dependence of spin-valley coupling in SiMOS and Si/SiGe, we perform free induction decay measurements with magnetic field oriented along a variety of directions defined with respect to the crystallographic axes shown in Fig.~\ref{fig:QubitOperation}b. We first initialize a well-separated $S$(3,1) state, such that $J$ is negligible, and allow it to evolve under rotations driven by $\Delta E_Z$. By varying the evolution time, we monitor the ST rotation frequency in order to probe the effective $\Delta E_Z$ generated by differences in $g$-factor and spin-valley coupling. We then repeat these free induction decay measurements while the magnetic field magnitude is continuously ramped over a fixed field orientation.

A representative free induction decay measurement on the Si/SiGe device is shown in Fig.~\ref{fig:QubitOperation}c. Coherent rotations between S and T$_0$ states are induced by the difference in Zeeman energy splitting between the two QDs. Here, S-T$_0$ decoherence is primarily driven by $^{29}$Si and $^{73}$Ge nuclear spins in the barriers and residual $^{29}$Si present in the purified quantum well, which flip-flop with each other when left unoccupied by electrons in the (4,0) state, and generate a slowly varying Overhauser field that drives noise on $\Delta E_\mathrm{Z}$, resulting in an inhomogeneous dephasing time of $T_{2}^{*} \approx 4.4$ $\mu$s. Fig. ~\ref{fig:QubitOperation}d shows repeated free induction decay measurements, also in Si/SiGe, while the external magnetic field magnitude is ramped along a fixed direction parallel to the crystallographic $[100]$ axis. Two discontinuities in the evolution frequency appear near $\sim$324 and $\sim$412 mT, identifying resonant spin-valley hot spots that occur in the two QDs occupied by the ST qubit.  

\begin{figure}[t]
	\centering
	\includegraphics[width=\linewidth]{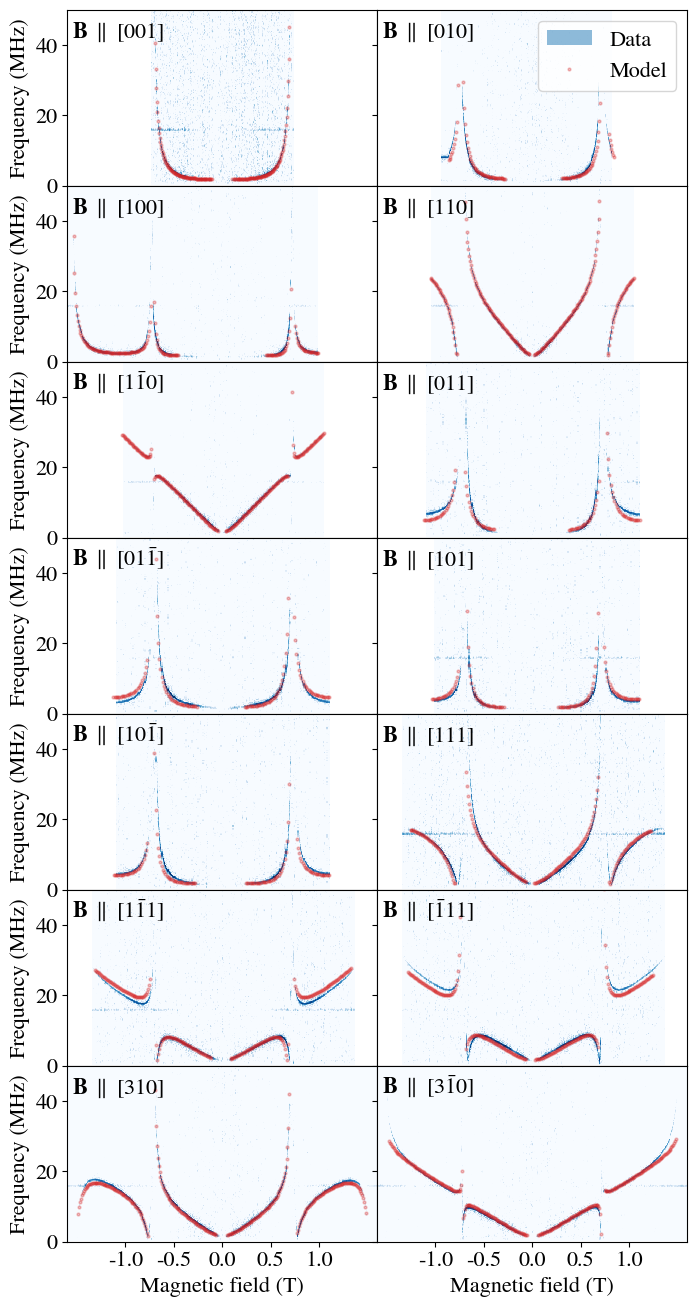}
	\caption{\textbf{SiMOS measurements} Measured ST rotation frequency derived from the FFT of repeated free induction decay measurements in SiMOS as $B$ magnitude is varied along fourteen orientations. The FFT data are normalized for plotting. The model fit is superimposed in red. Two divergences appear, corresponding to Zeeman resonances with the valley splittings in the two quantum dots.}
	\label{fig:ExperimentalMagnetospectroscopy_SiMOS}
\end{figure}

\begin{figure}[t]
	\centering
	\includegraphics[width=0.4\textwidth]{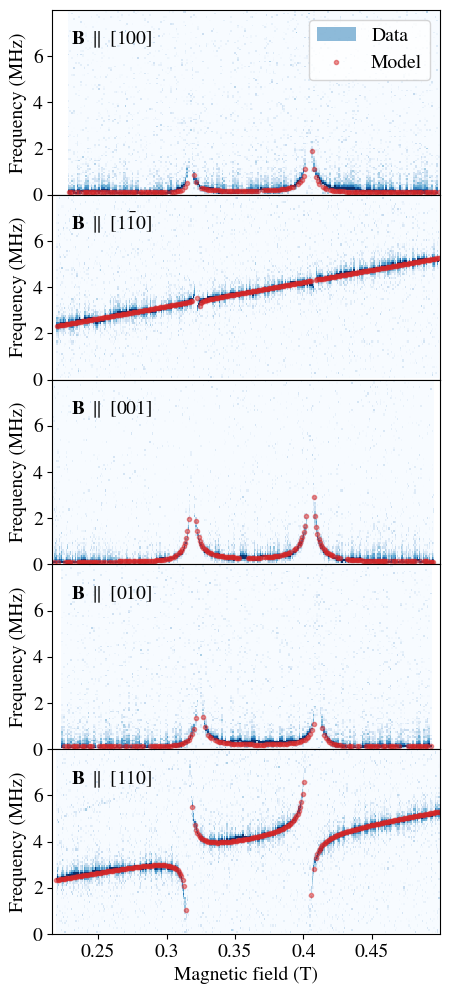}
	\caption{\textbf{Si/SiGe measurements} Measured ST rotation frequency derived from FFT of repeated free induction decay measurements in a Si/SiGe device along five $B$ orientations. The FFT data are normalized for plotting. The model fit is superimposed in red. Two divergences appear, corresponding to Zeeman resonances with the valley splittings in the two quantum dots. The data in the top panel are derived from the repeated free induction decay data in Fig.~\ref{fig:QubitOperation}d.}
	\label{fig:ExperimentalMagnetospectroscopy_SiSiGe}
\end{figure}

We acquired free induction decay measurements in SiMOS (Si/SiGe) over fourteen (five) selected magnetic field orientations overall, with their frequency components shown in Fig.~\ref{fig:ExperimentalMagnetospectroscopy_SiMOS} (Fig.~\ref{fig:ExperimentalMagnetospectroscopy_SiSiGe}). Spin-valley hot spots in each of the two QDs appear as divergences or discontinuities in ST rotation frequency when the valley splitting energy $\Delta_{\mathrm{vs},i}$ equals the Zeeman energy $E_{\mathrm{Z},i}=g_i\mu_\mathrm{B} B$ of each particular QD$_i$, where the index $i=$ A,B refers to either dot A or B, $g_i$ are the Landé $g$-factors, $\mu_\mathrm{B}$ is the Bohr magneton, and $B$ is the external magnetic field. We use the labeling convention here that dot A (B) refers to the QD having the lower (higher) valley splitting of the given pair. At the hot spot, the valley and spin degrees of freedom become mixed and the ST rotation frequency becomes a sensitive probe of the valley splitting energy. We note that without further information we cannot assign valley splitting values to one QD over another due to the S and T$_0$ states having equal support in both QDs. Additional measurements to characterize valley splitting variation as a function of voltages applied to gate electrodes may allow for unambiguously identifying to which QD the valley hot spots belong. Alternatively, as we detail later, measuring a different pair of QDs with one QD in common allows for such an identification.

\begin{table}[ht]
\begin{tabular}{| c | c | c |}
\hline
\textbf{Parameter} & \textbf{SiMOS} & \textbf{Si/SiGe} \\
\hline
Valley splitting, $\Delta_{\mathrm{vs,A}}$ ($\mathrm{\mu eV}$) & 83.1(9) & 36.81(1) \\
\hline
Valley splitting, $\Delta_{\mathrm{vs,B}}$ ($\mathrm{\mu eV}$) & 180.3(3) & 46.86(1) \\
\hline
Rashba SOC, $\Delta \alpha$ & -3(1)$\times 10^{-5}$& 4.6(5) $\times 10^{-6}$ \\
\hline
Dresselhaus SOC, $\Delta \beta$ & 2.04(1)$\times 10^{-3}$ & -7.585(5)$\times 10^{-4}$\\
\hline
Spin-valley coupling, $\gamma_{\mathrm{A}}$ ($\mathrm{\mu eV}$) & 0.730(3) & 0.0504(2) \\
\hline
Spin-valley coupling, $\gamma_{\mathrm{B}}$ ($\mathrm{\mu eV}$) & 0.87(2) & 0.0571(2) \\
\hline
Spin-valley phase, $\eta_{\mathrm{A}}$ (rad) & 0.553(8) & 0.615(7)\\
\hline
Spin-valley phase, $\eta_{\mathrm{B}}$ (rad) & 0.84(2) & 0.818(8)\\
\hline
\end{tabular}
\caption{\textbf{Fit valley splitting and SOC parameters for SiMOS and Si/SiGe devices.} Si/SiGe QDs measured here are (QD$_1$,QD$_2$). For the valley splitting of SiMOS QD$_A$, our reported uncertainty incorporates and is dominated by the observed dependence on magnetic field orientation shown in Fig. \ref{fig:Valley_splitting_angular_dependence}. The reported uncertainty for the valley splitting in SiMOS QD$_B$ corresponds to the magnetic field sweep orientations $\lbrace [100], [310] \rbrace$ in which the second hot spot is evident.}
\label{tab:results_table}
\end{table}

\begin{figure}[t]
	\centering
	\includegraphics[width=0.55\textwidth]{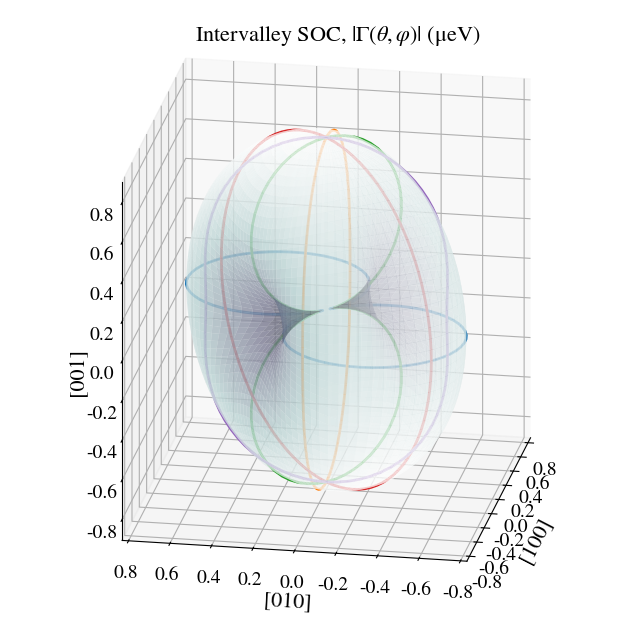}
	\caption{\textbf{Three-dimensional polar plot of the fit intervalley SOC strength as a function of magnetic field orientation for QD$_B$ of SiMOS device.} The five intersecting planes $(\theta,\varphi) \in $ $\lbrace (\pi/2,[0,2\pi])$, $([0,2\pi],0)$, $([0,2\pi],-\pi/4)$, $([0,2\pi],\pi/4)$, $([0,2\pi],\pi/2) \rbrace$ containing the probed B-field orientations are represented as intersecting curves. There is a clear node of the intervalley SOC that lies in the $x$-$y$ plane nearly along the $\mathrm{[1 \bar{1} 0]}$ axis. The polar plot for QD$_A$ and both dots of the Si/SiGe device have similar form given by Eq. \ref{eq:abs_SV_matrix_element}, with appropriate rescaling of the magnitude of the spin-valley coupling and slight adjustment of the node azimuthal angle.}
	\label{fig:IntervalleySOCPolarPlot}
\end{figure}

\begin{figure}[t]
	\centering
	\includegraphics[width=0.45\textwidth]{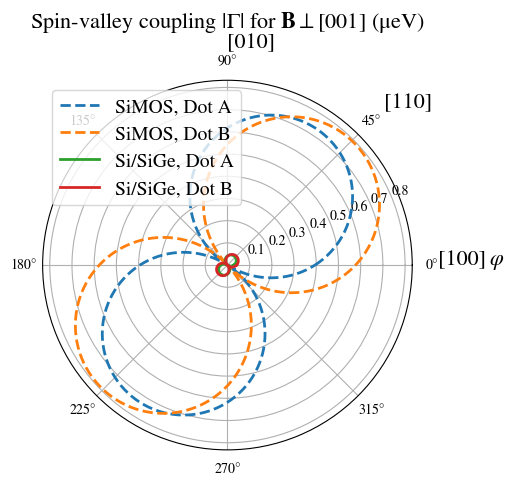}
	\caption{\textbf{Angular dependence of spin-valley coupling for $\mathbf{B} \perp [001]$.} The fit spin-valley coupling for both SiMOS and Si/SiGe devices has a maximum for magnetic field applied along the $[110]$ and $[\bar{1}\bar{1}0]$ axes. The magnitude of spin-valley coupling is an order of magnitude larger for SiMOS as compared with Si/SiGe.}
	\label{fig:Spin_valley_polar_plot}
\end{figure}

\begin{figure}[t]
	\centering
	\includegraphics[width=0.5\textwidth]{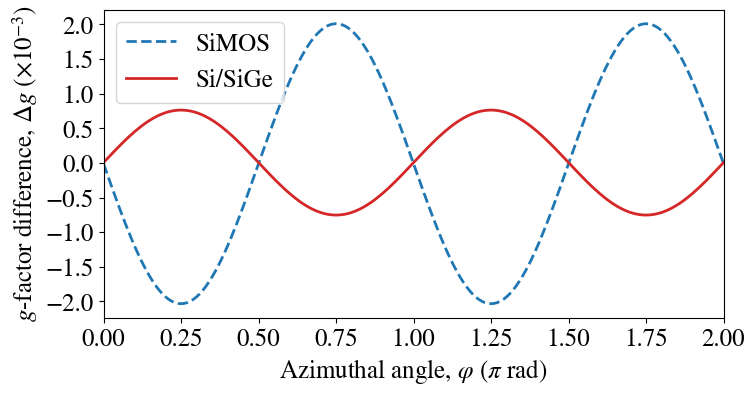}
	\caption{\textbf{Angular dependence of $g$-factor difference between QD$_A$ and QD$_B$ for $\mathbf{B} \perp [001]$.} The angular dependence for both devices is equivalent up to a $\pi/2$ rotation and rescaling due to the significantly larger Dresselhaus than Rashba SOC contributions in both cases.}
	\label{fig:delta_g_in-plane}
\end{figure}

\section{Discussion}
\label{sec:Discussion}
The model we use to fit to the data in Figs.~\ref{fig:ExperimentalMagnetospectroscopy_SiMOS} and \ref{fig:ExperimentalMagnetospectroscopy_SiSiGe} is captured by the four-level Hamiltonian that we now describe\cite{Jock2022,Volmer2024,Tomic2025}. The model consists of expanding the intravalley spin-orbit Hamiltonian for a ST qubit located in QD$_A$ and QD$_B$, written in the ground spin-valley eigenbasis (subscript 0) spanning $\lbrace \vert \tilde{\uparrow}_{0}^{A} \tilde{\downarrow}_{0}^{B} \rangle, \vert \tilde{\downarrow}_{0}^{A} \tilde{\uparrow}_{0}^{B} \rangle \rbrace$ to include only the interacting first excited valley states (subscript 1) spanning $\lbrace \vert \tilde{\downarrow}_{1}^{A} \tilde{\downarrow}_{0}^{B} \rangle, \vert \tilde{\downarrow}_{0}^{A} \tilde{\downarrow}_{1}^{B} \rangle \rbrace$.  The tilde ( $\tilde{}$ ) notation denotes that the spin states are defined relative to $\mathbf{B}(\theta,\phi)$. In the model Hamiltonian,

\begin{equation}
    H \!= \! \left( \begin{array}{cccc}
\delta B/2 & J/2 & \Gamma_{A} & 0 \\
J/2 & -\delta B/2 & 0 & \Gamma_{B} \\
\Gamma_{A}^{*} & 0 & \Delta_{\mathrm{vs},A} \! - \! g\mu_{B} B & 0 \\
0 & \Gamma_{B}^{*} & 0 & \Delta_{\mathrm{vs},B} \! - \! g \mu_{B} B
\end{array}\right),
\label{eq:Four-level_Hamiltonian}
\end{equation}
the spin-valley couplings are $\Gamma_i$ and the difference in $g$-factors of the QDs are contained in $\delta=\delta(\theta,\phi)=\mu_B(\Delta\alpha - \Delta\beta\sin(2\phi))\sin^2(\theta)$ \cite{Jock2018}. The differences in Rashba and Dresselhaus intervalley SOC between the QDs are quantified by $\Delta\alpha$ and $\Delta\beta$, respectively. The magnetic field angular dependence of the spin-valley coupling is given by 
\begin{equation}
\label{eq:abs_SV_matrix_element}
\vert \Gamma(\theta,\varphi) \vert = \frac{\gamma}{2} \sqrt{3 \! + \! \cos(2\theta) \! - \! 2 \cos(2(\varphi \! + \! \eta)) \sin^{2}(\theta)},
\end{equation}
where $\gamma$ is the magnitude of the relevant intravalley spin-orbit matrix elements and $\eta$ is the phase orientation of the maximum of $\vert \Gamma(\theta,\varphi) \vert$ relative to the [100], [010] crystallographic axes. We provide further details of the four-level model and how it maps to ST rotation frequency in Appendix~\ref{sec:Model}. 

The resulting fit parameters for SiMOS and Si/SiGe devices are summarized in Table~\ref{tab:results_table}, with details of our statistical analysis described in Appendix~\ref{appendix:Fits_and_statistics}. Our measured valley splitting energies in the Si/SiGe device are consistent with previously reported measurements in the literature \cite{Mi2017,PaqueletWuetz2022,Volmer2024,Marcks2025}. 
Both valley splittings inferred from the SiMOS hot spots are larger than those in Si/SiGe by a factor of two to five, with a correspondingly larger difference between the two. Similarly, the spin-valley coupling magnitude $\gamma$ for each hot spot is an order of magnitude larger in SiMOS than in Si/SiGe, with a correspondingly larger difference between the same-device QD pairs.

Our results appear to be largely consistent with previously-reported spin-valley coupling values in the literature. Similar measurements of a Si/SiGe double QD device with a 10 nm thick strained Si quantum well observed spin-valley couplings, $g$-factor difference, and valley splitting values that are highly consistent with our measurements~\cite{Volmer2024}. Measurements of a Si/SiGe double QD making use of the S/T$_-$ encoding observed spin-valley coupling of a smaller magnitude~\cite{Cai2023}. Measurements of spin-valley hot spots in SiMOS appear to be of a similar order of magnitude to what we observed \cite{Yang2013,Hao2014}, though other magnetic field angular-dependence measurements of spin-valley coupling in SiMOS are an order of magnitude smaller than what we have observed~\cite{Zhang2020}. Moreover, measurements in CMOS silicon nanowire devices show comparable spin-valley couplings larger\cite{Corna2018} than or of a similar magnitude\cite{Spence2022} to what we have measured.

The full angular dependence of $\vert\Gamma(\theta,\phi)\vert$  for QD$_B$ in the SiMOS device is shown in Fig.~\ref{fig:IntervalleySOCPolarPlot}, where the colored rings outline the planes over which the data were acquired. The angular dependence for the other measured QDs is similar, differing only be a rescaling and slight rotation of the azimuthal angle of the node. A cross-section of the three-dimensional polar plot is shown in Fig.~\ref{fig:Spin_valley_polar_plot}. The spin-valley phase offset $\eta\approx\pi/4$ (extracted fit parameters shown in Table~\ref{tab:results_table}) in both cases of dot pairs is apparent from the alignment of the lobes of maximum $\vert \Gamma \vert$ with the $[110]$ and $[\bar{1}\bar{1}0]$ crystallographic axes. 
We suspect that slight variation in $\eta$ between QDs on the same device may be related to dot-to-dot differences in sampled interfacial disorder and/or quantum dot confinement potentials, though further modeling would be needed to assess sources of variability.

The fit results show Rashba and Dresselhaus SOC parameters to be slightly larger in SiMOS than in Si/SiGe. The resulting azimuthal angular dependence of the \textit{difference} in $g$-factors (Fig.~\ref{fig:delta_g_in-plane}) shows that the intravalley SOC is distinctly maximized at $\phi=(2n - 1)\pi/4$, with a $\pi/2$ phase offset between SiMOS and Si/SiGe devices. The inverted sign of $\Delta g$ can be understood simply from the magnetic field gradient pointing in opposite directions with respect to each device's designated QD with larger $\Delta_\mathrm{vs}$.
This highlights the distinguishing behavior of spin-valley coupling and intravalley SOC affecting the ST qubit. While both effects stem from the same interfacial SOC mechanisms, the former is sensitive to the absolute value localized at each QD, while the latter is sensitive only to differences between them.

\begin{figure*}[t]
	\centering
	\includegraphics[width=0.8\linewidth]{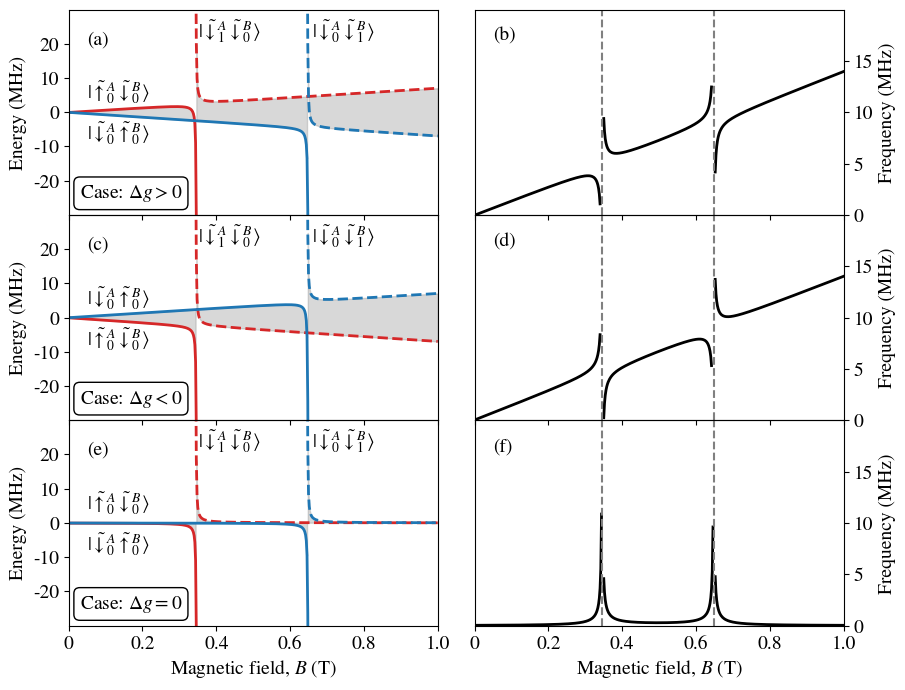}
	\caption{\textbf{Taxonomy of spin-valley hotspots in a double quantum dot.} Energy levels for two electron spins in a pair of quantum dots encoding a ST qubit. In the left column, we plot the energy levels of the relevant four spin-valley-orbital states of the system as a function of applied magnetic field for three different cases of $g$-factor differences (a) $\Delta g > 0$, (c) $\Delta g < 0$, and (e) $\Delta g = 0$. The energy difference corresponding to singlet-triplet rotations is indicated as the pairs of lines connected by the shaded gray regions. The solid and dashed curves of same color correspond to the pair of states interacting via spin-valley coupling, with red (blue) corresponding to the spin-valley hotspot in quantum dot A (B), respectively. In the right column, we plot the corresponding singlet-triplet rotation frequencies as a function of magnetic field. The magnetic field at which Zeeman splitting equals valley splitting in dot A (B) are indicated by the left (right) vertical dashed lines. Based on the sign and magnitude of the $g$-factor differences, we observe three distinct frequency versus field dependencies. For the illustrative example shown here, we assume $\Delta g \in \lbrace 10^{-3},-10^{-3},0 \rbrace$, $\Delta_{\mathrm{vs},A}=40 \ \mathrm{\mu eV}$, $\Delta_{\mathrm{vs},B}=75 \ \mathrm{\mu eV}$, and $\vert \Gamma_{A} \vert = \vert \Gamma_{B} \vert = 0.1 \ \mathrm{\mu eV}$.}
	\label{fig:EnergyDiagramTaxonomy}
\end{figure*}

To understand further the interplay of $g$-factor differences and spin-valley coupling in the observed ST rotation frequencies as a function of magnetic field, we organize the field dependence into three qualitative forms. As shown in Fig \ref{fig:EnergyDiagramTaxonomy}, these stem from the cases: (1) significant SOC drives an effective field gradient of sign $\Delta E_{\mathrm{Z}} > 0$, (2) significant SOC drives an effective field gradient of sign $\Delta E_{\mathrm{Z}} < 0$, or (3) the $g$-factor difference due to intravalley SOC is negligible at all fields relative to intervalley SOC, resulting in rotation frequency divergences at each valley splitting. The first two cases correspond to what are qualitatively oblique asymptotes with opposite handedness for each valley splitting resonance. The width of the divergence relates to the strength of the spin-valley coupling, while the overall slope of the frequency dependence on magnetic field strength indicates the magnitude of the $g$-factor difference between QDs.

The origin of these functional forms can be understood by considering the three states that are most relevant around the hot spot. If we are considering the hot spot associated with the valley splitting of QD$_B$, then these three states are $\ket{\tilde{\uparrow}_{0}^{A} \tilde{\downarrow}_{0}^{B}}$, $\ket{\tilde{\downarrow}_{0}^{A} \tilde{\uparrow}_{0}^{B}}$, $\ket{\tilde{\downarrow}_{0}^{A} \tilde{\downarrow}_{1}^{B}}$.  Since there is no intervalley SOC coupling between $\ket{\tilde{\downarrow}_{0}^{B}}$ and $\ket{\tilde{\downarrow}_{1}^{B}}$ due to the spin states being the same, the $\ket{\tilde{\uparrow}_{0}^{A} \tilde{\downarrow}_{0}^{B}}$ state is effectively ``inert'' with respect to QD$_B$'s spin-valley hot spot. Consequently, the energy of this state has a simple linear slope due to the effective magnetic field gradient arising from the $g$-factor difference between dots, through intravalley SOC. However, the other two states $\ket{\tilde{\downarrow}_{0}^{A} \tilde{\uparrow}_{0}^{B}}$ and $\ket{\tilde{\downarrow}_{0}^{A} \tilde{\downarrow}_{1}^{B}}$ do interact through intervalley SOC, producing a corresponding anticrossing at the hot spot. The ST qubit evolution frequency is given by the energy difference between the eigenstates closest to the span of $\ket{\tilde{\uparrow}_{0}^{A} \tilde{\downarrow}_{0}^{B}}$ and $\ket{\tilde{\downarrow}_{0}^{A} \tilde{\uparrow}_{0}^{B}}$, giving rise to the evolution frequencies shown in Fig. \ref{fig:EnergyDiagramTaxonomy}.

\begin{figure}[!ht]
	\centering
	\includegraphics[width=\linewidth]{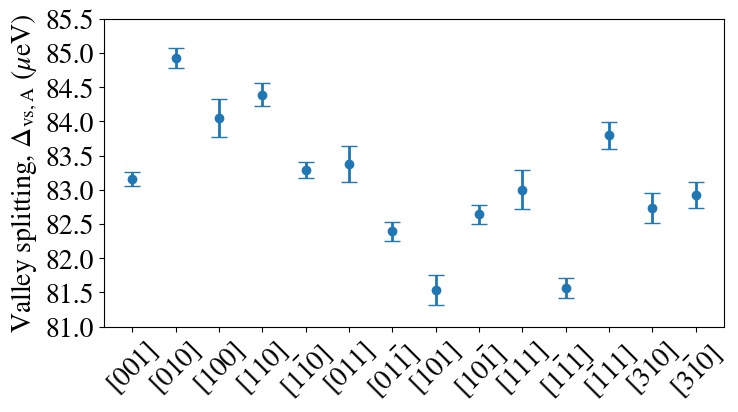}
	\caption{\textbf{Magnetic field orientation dependence of valley splitting for SiMOS QD$_A$.} Our estimate of $\Delta_{\mathrm{vs,A}}$ is based on shifts of the hot spot critical field for QD$_A$. The error bars denote 95\% confidence intervals.}
	\label{fig:Valley_splitting_angular_dependence}
\end{figure}

In fitting our model to the data, we find that for SiMOS the position of the spin-valley hot spot critical fields varies significantly relative to those of the measured Si/SiGe device. Our interpretation of this, and the one we implement in the model fitting, is that the valley splitting may have a measurable magnetic field dependence~\cite{Friesen2006, Cai2023}. Considering that the magnetic fields probed in the SiMOS measurements are significantly larger than for Si/SiGe, we suspect that the correspondingly larger magnetic confinement may be driving a magnetic field-dependent valley splitting. Since the valley splitting depends considerably on details of the interface, modified quantum dot confinement resulting from rotated magnetic field orientation may lead to slightly different sampling of the interface and, consequently, shifted valley splitting. Based on observed variation of hot spot location as a function of magnetic field orientation, the estimated angular dependence of the valley splitting in SiMOS device QD$_A$ is shown in Fig. \ref{fig:Valley_splitting_angular_dependence}.

We suspect that the stronger spin-valley coupling for the SiMOS device may stem from the enhanced confinement against the Si/SiO$_2$ interface typical of SiMOS quantum dots. For a Si/SiGe quantum well, on the other hand, the lower and upper SiGe provides the confinement potential along the growth axis, with the smaller band offset between Si/SiGe as compared with Si/SiO$_2$ somewhat limiting the strength of electric field that may be imposed before undesirable tunneling out of the quantum well occurs. As for valley splitting\cite{Pena2024}, we anticipate that thinner quantum wells may exhibit enhanced spin-valley coupling, given the greater overlap with the Si/SiGe interface that also enhances valley splitting. Similarly for Si/SiGe quantum wells with nonzero Ge concentration within the well, theoretical work has predicted that SOC may also be enhanced \cite{Woods2023}.

\begin{figure}[!ht]
	\centering
	\includegraphics[width=0.7\linewidth]{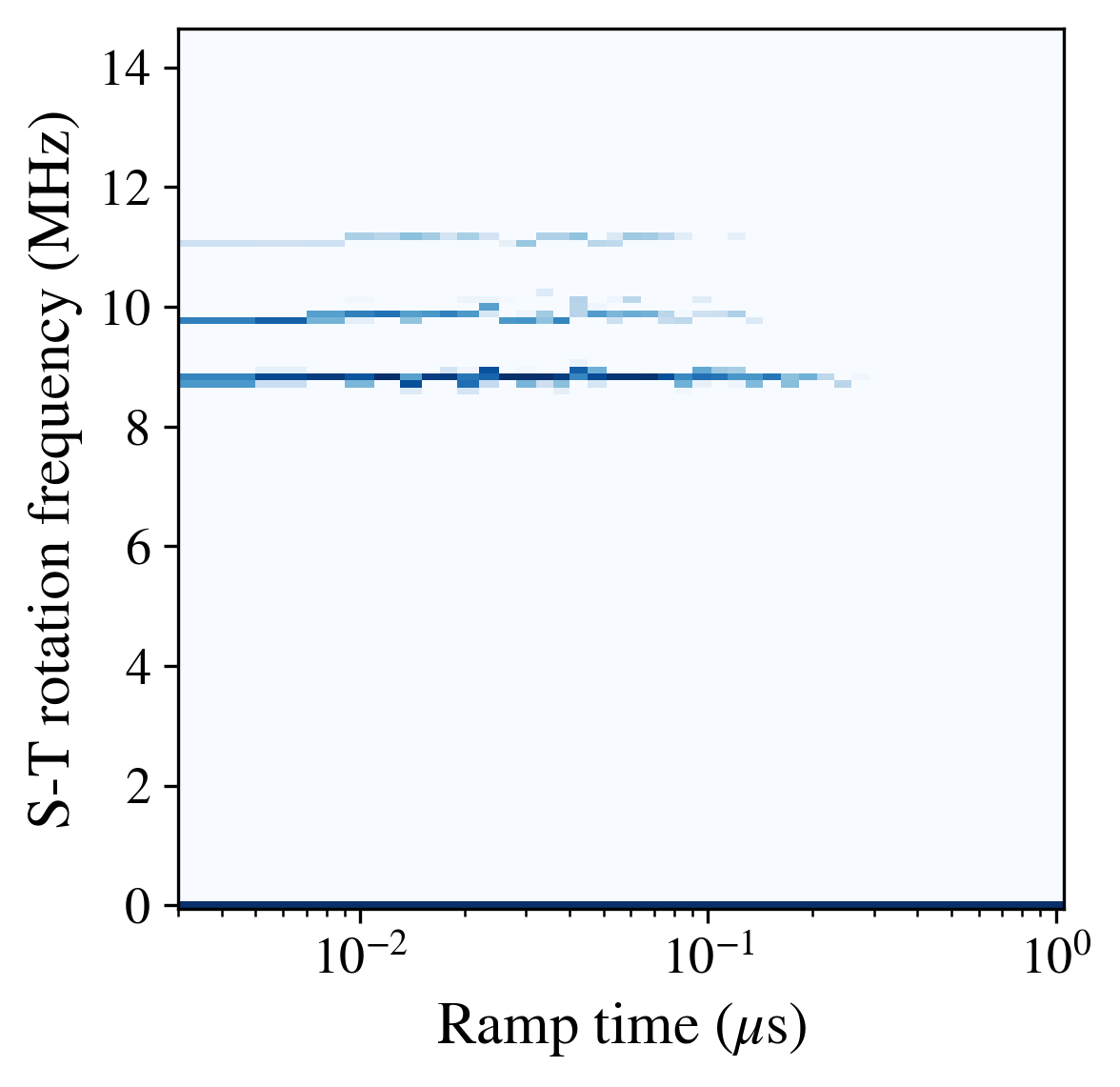}
	\caption{\textbf{Excited valley state population.} Frequency components of free induction decay measurements versus ramp time from S(4,0) to S(3,1). The color axis is the relative amplitude of the FFT in evolution time. For this measurement, $\textbf{B}~\vert\vert~[110] =$ 523 mT in the (QD2,QD3) configuration in the Si/SiGe device.}
	\label{fig:ExcitedValleys}
\end{figure}

We performed a similar set of measurements for a ST qubit occupying another pair of QDs (QD$_2$,QD$_3$) in the Si/SiGe triple QD, finding multiple frequency components of the free induction decay. Similar observations have been reported for a SiMOS device having valley splittings equivalent to those of our Si/SiGe device~\cite{Tomic2025}. These results are described in Appendix \ref{appendix:Additional_measurements}, with our estimates for valley splitting and spin-valley coupling shown in Table \ref{tab:multi_frequencyresults_table}. We find that our observation of multiple frequency components is consistent with population of excited valley eigenstates, based on an analysis of frequency component amplitude as a function of the ramping time for state preparation shown in Fig. \ref{fig:ExcitedValleys}. The valley splitting and spin-valley coupling of QD$_A$ in the measurement of (QD$_2$,QD$_3$) being similar to that of QD$_B$ in the earlier measurement of (QD$_1$,QD$_2$), we suspect that the middle QD likely corresponds to QD$_B$ and QD$_A$ of the (QD$_1$,QD$_2$) and (QD$_2$,QD$_3$) measurements, respectively.

\section{Conclusion}
\label{sec:Conclusion}
In this work, we have measured SOC-driven effective magnetic field gradients in a singlet-triplet qubit as a function of applied magnetic field strength and orientation. We observe magnetic field magnitude dependence characteristic of resonances between Zeeman and valley splitting energy corresponding to spin-valley hot spots. We formulated a simple model that captures the full angular dependence of intra- and intervalley SOC. We performed these measurements for two material systems, SiMOS and Si/SiGe, finding similarities and differences between the two. Observed valley splitting in the SiMOS device is significantly higher than for the Si/SiGe device, while the observed spin-valley coupling in SiMOS is an order of magnitude larger than for Si/SiGe. On the other hand, the angular dependence of spin-valley coupling is nearly the same for both material systems, exhibiting maxima for all four quantum dots measured along the [110] crystallographic axes. Moreover, the $g$-factor differences between quantum dots are of a similar magnitude for both material systems.

A simple way to minimize the effective magnetic field gradient due to SOC from intravalley SOC-induced $g$-factor differences is to apply the magnetic field normal to the interface ($\mathbf{B} \parallel [001]$). However, when operating the ST qubit with magnetic fields not far from the hot spot, the effect of the spin-valley hot spot on qubit dynamics can be significant and is nearly maximal for $\mathbf{B}$ normal to the interface. In this case it may also be useful to consider the hot spot linewidths when making the choice of field direction.  Fig.~\ref{fig:Linewidths_SiMOS} (Fig.~\ref{fig:Linewidths_SiSiGe}) shows the same FFT data for the SiMOS (Si/SiGe) device shown in Fig.~\ref{fig:ExperimentalMagnetospectroscopy_SiMOS} (Fig.~\ref{fig:ExperimentalMagnetospectroscopy_SiSiGe}) overlaid now with linewidths of the FFTs on the secondary $y$-axis. In general, the linewidth grows with proximity to the hot spot, reaching values in excess of 5 MHz, which we believe to be primarily dephasing due to enhanced charge noise sensitivity\cite{Jock2022}. However, the linewidth broadening is more localized in $B$ along specific field orientations (\textit{e.g.}, $[1\bar{1}0]$ or $[3\bar{1}0]$), which could provide a useful choice of magnetic field orientation when attempting to avoid enhanced dephasing due to spin-valley hot spots. 

Ultimately, a qubit's limiting error sources will drive the choice for magnetic field magnitude~\cite{Petit2018,Borjans2019}. Our measurements and analysis provide insight into magnetic field orientation dependence, where the role of SOC and spin-valley characteristics should be carefully considered in the context of qubit operation. For example, a qubit in Si/SiGe that is more limited by magnetic noise than spin relaxation or charge noise could benefit from a field orientation that intentionally minimizes $\Delta g$, such as $[001]$. On the other hand, operating a qubit in SiMOS that may be more exposed to spin relaxation or charge noise would be better insulated from $T_1$ or dephasing mechanisms when the magnetic field is oriented along $[1\bar{1}0]$, where linewidth of the ST frequency is narrower near the hot spots. In the future, it would be interesting to compare these measured spin-valley couplings and $g$-factor differences with detailed atomistic models of SiMOS or Si/SiGe interfaces.

\textbf{Acknowledgements-}
Sandia National Laboratories is a multi-mission laboratory managed and operated by National Technology and Engineering Solutions of Sandia, LLC., a wholly owned subsidiary of Honeywell International, Inc., for the U.S. Department of Energy's National Nuclear Security Administration under contract DE-NA-0003525.
Research was sponsored in part by the Army Research Office and was accomplished under Cooperative Agreement Number W911NF-22-2-0037. The views and conclusions contained in this document are those of the authors and should not be interpreted as representing the official policies, either expressed or implied, of the Army Research Office or the U.S. Government. The U.S. Government is authorized to reproduce and distribute reprints for Government purposes notwithstanding any copyright notation herein.
We acknowledge support from Intel Corporation for providing the Si/SiGe device.

\textbf{Author contributions-}
R.M.J., N.D.F., and N.T.J. designed the experiments and analyzed the results. R.M.J. and N.D.F. performed the central measurements presented in this work. N.T.J. developed the theoretical description of the results and performed the statistical analysis. M.R. performed supporting measurements on a similar device that established repeatability of observations. A.M.M. contributed to the development of the theoretical model. N.T.J., R.M.J., N.D.F., M.R., A.M.M., A.D.B., M.S.C., and D.R.L. discussed central results throughout the project. For the SiMOS devices, D.R.W. and M.S.C. designed the process flow, fabricated the devices, and designed/characterized the $^{28}$Si material growth for this work. M.S.C., D.R.L., and R.M.J. supervised the combined effort, including coordinating fabrication and identifying modeling needs for the experimental path. N.T.J., N.D.F., and R.M.J. wrote the manuscript with input from all co-authors.

\appendix
\section{Experimental setup}
\label{sec:measurement}
We performed our measurements in separate $^3$He/$^4$He dilution refrigerators (SiMOS in an Oxford Triton and Si/SiGe in a Bluefors LD400) with a base temperature around 8-10 mK. The effective electron temperatures in the devices were about 150 and 250 mK for SiMOS and Si/SiGe experiments, respectively. Fast pulses to the quantum dot gates are applied through cryogenic RC bias-Ts. The external magnetic fields were applied using a 3-axis vector magnet. We performed cryogenic preamplification of the SET current for charge sensing using a heterojunction bipolar transistor (HBT) \cite{Curry2015,Curry2019}. The current was converted to voltage using a transimpedance amplifier (Femto DHPCA-200+), demodulated on a lock-in amplifier (Zurich Instruments), and digitized on an oscilloscope (NI5110).

Superconducting magnets are known to have a nontrivial degree of hysteresis when reversing the field ramp direction. For our measurements of the Si/SiGe device, we estimate this effect to introduce an uncertainty of about 2-3 mT in magnetic field magnitude. Our magnetic field sweeps for measurements of the SiMOS device were symmetric through zero, allowing for this hysteretic effect to be compensated. The estimated hysteresis values for the SiMOS measurements are shown in Table \ref{tab:BOffsets}. 

\begin{table}
	\centering
	\begin{tabular}{| c | c |}
		\hline
		\textbf{Orientation} & $B_{\mathrm{offset}}$ (mT) \\ \hline
		$[10 \bar{1}]$ & -13.3 \\ \hline
		$[0 1 \bar{1}]$ & -16.8 \\ \hline
		$[1 \bar{1}0]$ & -6.0 \\ \hline
		$[3 \bar{1} 0]$ & 4.6 \\ \hline
		$[100]$ & 2.0 \\ \hline
		$[310]$ & -5.0 \\ \hline
		$[110]$ & -3.2 \\ \hline
		$[010]$ &  14.9 \\ \hline
		$[\bar{1} 1 1]$ & -15.6 \\ \hline
		$[1 \bar{1} 1]$ & -14.7 \\ \hline
		$[101]$ & -13.1 \\ \hline
		$[111]$ & -23.6 \\ \hline
		$[011]$ & -8.0 \\ \hline
		$[001]$ & -1.9 \\
		\hline
	\end{tabular}
	\caption{\label{tab:BOffsets} \textbf{Magnetic field offsets for each sweep direction for SiMOS measurements.} We attribute these offsets to synchronization error of measurement and hysteresis. Given the initial experimental estimate for the (signed) magnetic field strength $B$, the offset correction takes $B \to B + B_{\mathrm{offset}}$.}
\end{table}

Additionally, the runtime of a single free induction decay linescan may limit the magnetic field resolution. For high contrast oscillations, each free induction decay scan over variable wait time was averaged over 100 shots taking about 5-6 seconds to complete, while the magnetic field was ramped at a rate of 10 mT/minute or $\sim$0.2 mT/s. This equals a magnetic field uncertainty of up to 1 mT, corresponding to about 0.1 $\mu$eV in energy.

\section{Model}
\label{sec:Model}
The following model follows closely the analysis previously detailed in \cite{Jock2022}. To avoid ambiguity when considering arbitrary B-field orientations, we define the $\ket{\uparrow}$ and $\ket{\downarrow}$ states as the spin states relative to the crystallographic [001] axis, which is the normal to the Si/SiO$_2$ or Si/SiGe interface. We use a tilde notation (e.g. $\ket{\tilde{\uparrow}}$) to indicate a spin state defined relative to the quantization axis specified by a given $\mathbf{B}$ orientation.

Let the two-electron spin states for the regime of well-separated electrons be given by the following. Note that the charge configuration we assume from now on is either (1,1) or (1,3) electron occupancy, with the latter treating the ground valley eigenstate in the second dot as a filled shell.

We parameterize the applied magnetic field relative to the intrinsic crystal axes [100], [010], [001] as
\begin{eqnarray}
\mathbf{B} & = & (B_{x}, B_{y}, B_{z}) \nonumber \\
& = & B \big( \sin \theta \cos \varphi, \sin \theta \sin \varphi, \cos \theta \big),
\end{eqnarray}
with $\theta$ corresponding to an elevation angle indicating deviation from [001], the normal to the Si/SiO$_2$ or SiGe/Si/SiGe quantum well, and $\varphi$ the azimuthal angle away from the [100] crystallographic axis. The spin eigenstates relative to this magnetic field orientation are
\begin{eqnarray}
    \vert \tilde{\uparrow} \rangle & = & e^{-i \varphi/2} \cos(\theta/2) \vert \! \uparrow \rangle + e^{i \varphi/2} \sin(\theta/2) \vert \! \downarrow \rangle \\
    \vert \tilde{\downarrow} \rangle & = & -e^{-i \varphi/2} \sin(\theta/2) \vert \! \uparrow \rangle + e^{i \varphi/2} \cos(\theta/2) \vert \! \downarrow \rangle.
\end{eqnarray}

\subsection{Intravalley spin-orbit coupling}
We assume the spin-orbit coupling Hamiltonian in a given quantum dot to be a sum of Rashba and Dresselhauss terms $H_{\mathrm{SO}} = H_{\mathrm{R}} + H_{\mathrm{D}}$, with
\begin{eqnarray}
H_{\mathrm{R}} & = & \alpha \left( P_{y} \sigma_{x} - P_{x} \sigma_{y}\right) \\
H_{\mathrm{D}} & = & \beta \left( P_{x} \sigma_{x} - P_{y} \sigma_{y}\right),
\end{eqnarray}
where $P_{x,y}$ are kinetic momenta along the $x$ and $y$ crystallographic axes and the factors $\alpha$ and $\beta$ potentially have a spatial dependence to account for interfacial SOC effects \cite{Golub2004,Prada2011,Jock2018,Ruskov2018,Ferdous2018,Hosseinkhani2021}. Note that the resulting Rashba or Dresselhaus coupling acting on the spin degrees of freedom will depend not only on the ``microscopic'' spin-orbit fields represented by $\gamma_{R}$ and $\gamma_{D}$, intrinsic to the interface, but also the degree of vertical confinement, quantum dot lateral extent, relative phase between $\pm z$ valleys of the ground and excited valley eigenstates, and any dipole matrix element between those valley eigenstates arising from disorder at the Si/SiO$_2$ or Si/SiGe interfaces \cite{Ruskov2018}. Note that the SOC Hamiltonian has no diagonal terms relative to the $\sigma_z$ spin eigenstates.

\subsection{Spin-valley coupling}
For a given quantum dot, let the ground and first-excited valley-orbital eigenstates be given by $\vert E_{0}\rangle$, $\vert E_{1}\rangle$, respectively. Adding in the spin degree of freedom, we are interested in evaluating the spin-orbit coupling between ground and first-excited valley-orbital states having opposite spin orientation. Denote 
\begin{equation}
\gamma_{\mathrm{R,D}}^{\uparrow \downarrow} = \langle E_{0} \uparrow \vert H_{\mathrm{R,D}} \vert E_{1} \downarrow \rangle,
\end{equation}
with $\langle E_{0} \vert \alpha P_{x,y} \vert E_{1} \rangle = \lambda^{\alpha}_{x,y}$ and $\langle E_{0} \vert \beta P_{x,y} \vert E_{1} \rangle = \lambda^{\beta}_{x,y}$.
We have
\begin{eqnarray}
\gamma_{\mathrm{R}}^{\uparrow \downarrow} & = & \lambda_{y}^{\alpha} + i \lambda_{x}^{\alpha} \\
\gamma_{\mathrm{R}}^{\downarrow \uparrow} & = & \lambda_{y}^{\alpha} - i \lambda_{x}^{\alpha} \\
\gamma_{\mathrm{D}}^{\uparrow \downarrow} & = & \lambda_{x}^{\beta} + i \lambda_{y}^{\beta} \\
\gamma_{\mathrm{D}}^{\downarrow \uparrow} & = & \lambda_{x}^{\beta} - i \lambda_{y}^{\beta}.
\end{eqnarray}
In the following, we furthermore assume that a relative global phase between $\vert E_{0} \rangle$ and $\vert E_{1} \rangle$ may be chosen such that the matrix elements $\lambda_{x,y}^{\alpha,\beta}$ are real. Making this assumption, we have $\gamma_{\mathrm{R},\mathrm{D}}^{\uparrow \downarrow} = (\gamma_{\mathrm{R},\mathrm{D}}^{\downarrow \uparrow})^{*}$. Denoting $\gamma_{\mathrm{R}}^{\uparrow \downarrow} + \gamma_{\mathrm{D}}^{\uparrow \downarrow} = \gamma e^{i \eta}$, we have
\begin{eqnarray}
\langle E_{0} \tilde{\uparrow} \vert H_{\mathrm{SO}} \vert E_{1} \tilde{\downarrow}\rangle & = & e^{i \varphi} \cos^{2} (\theta/2) \langle E_{0} \uparrow \vert H_{\mathrm{SO}} \vert E_{1} \downarrow \rangle \\
& & - e^{-i \varphi} \sin^{2} (\theta/2) \langle E_{0} \downarrow \vert H_{\mathrm{SO}} \vert E_{1} \uparrow \rangle \nonumber \\
& = & e^{i \varphi} \cos^{2} (\theta/2) \left( \gamma_{\mathrm{R}}^{\uparrow \downarrow} + \gamma_{\mathrm{D}}^{\uparrow \downarrow}\right) \nonumber \\
& & - e^{-i \varphi} \sin^{2} (\theta/2) \left( \gamma_{\mathrm{R}}^{\downarrow \uparrow} + \gamma_{\mathrm{D}}^{\downarrow \uparrow} \right) \nonumber \\
& = & \gamma \left( e^{i (\varphi + \eta)} \cos^{2} \frac{\theta}{2} - e^{-i (\varphi+\eta)} \sin^{2} \frac{\theta}{2} \right) \nonumber \\
& \equiv & \Gamma(\theta,\varphi) \label{eq:SV_mtx_element}
\end{eqnarray}
Hence, the magnetic field angular dependence of the magnitude of this matrix element is given by Eq. \ref{eq:abs_SV_matrix_element}.
As depicted in Fig. \ref{fig:IntervalleySOCPolarPlot}, this quantity takes its maximum of $\vert \gamma \vert$ on the great circle $\varphi \in \lbrace \pi/2 - \eta, 3\pi/2-\eta \rbrace$, which includes the case of magnetic field oriented normal to the plane of the quantum well. The phase $\eta$ dictates the orientation of this maximum relative to the $[100]$, $[010]$ crystallographic axes.

\subsection{Double quantum dot Hamiltonian}
We now consider the energy levels of the singlet-triplet qubit encoded within the double quantum dot system. We are interested in the dynamics of the singlet and unpolarized triplet states, $\vert S \rangle$ and $\vert T_{0} \rangle$, in the ground valley-orbital sector of the double quantum dot. These states are spanned by $\lbrace \vert \tilde{\uparrow}_{0}^{A} \tilde{\downarrow}_{0}^{B} \rangle, \vert \tilde{\downarrow}_{0}^{A} \tilde{\uparrow}_{0}^{B} \rangle \rbrace$, where for compactness we denote $\vert \tilde{\uparrow}_{0}^{A} \rangle = \vert E_{0}^{A} \rangle \vert \tilde{\uparrow} \rangle$, for example. For our considerations here, we assume that the exchange coupling $J$ is negligible. The splitting between these two states is set by the difference in $g$-factor between the dots \cite{Jock2018}, which has a magnetic field angular dependence determined by the spin-orbit coupling in each quantum dot. We are interested in coupling to excited valley states via the spin-valley coupling, so we furthermore include two additional polarized states given by $\vert \tilde{\downarrow}_{1}^{A} \tilde{\downarrow}_{0}^{B}\rangle$ and $\vert \tilde{\downarrow}_{0}^{A}\tilde{\downarrow}_{1}^{B} \rangle$, where the index 1 denotes the excited valley-orbital states $\vert E_{1}^{A,B} \rangle$. Within the four-dimensional space $\lbrace \vert \tilde{\uparrow}_{0}^{A} \tilde{\downarrow}_{0}^{B} \rangle, \vert \tilde{\downarrow}_{0}^{A} \tilde{\uparrow}_{0}^{B} \rangle, \vert \tilde{\downarrow}_{1}^{A} \tilde{\downarrow}_{0}^{B}\rangle, \vert \tilde{\downarrow}_{0}^{A}\tilde{\downarrow}_{1}^{B} \rangle \rbrace$, we have the Hamiltonian of Eq. \ref{eq:Four-level_Hamiltonian}, where $\Delta_{\mathrm{vs},A}$ and $\Delta_{\mathrm{vs},B}$ denote the valley splitting of dots A and B, respectively. Any residual exchange in the (1,1) charge configuration is given by $J$. The difference in $g$-factors in the dots is represented by $\Delta g(\theta,\varphi) = g_{A}(\theta,\varphi) - g_{B}(\theta,\varphi) = \delta(\theta,\varphi)/\mu_{B}$, with $\delta(\theta,\varphi)$ having the form \cite{Jock2018}
\begin{equation}
    \label{eq:delta_g_angular_dependence}
    \delta(\theta,\varphi) = \mu_{B}\left(\Delta \alpha - \Delta \beta \sin(2\varphi) \right)\sin^{2}(\theta),
\end{equation}
where $\Delta \alpha$ and $\Delta \beta$ are unitless factors quantifying the differences in Rashba and Dresselhaus SOC between quantum dots, respectively, and $\mu_{B} = 57.88 \ \mathrm{\mu eV/T}$ is the Bohr magneton.

The respective spin-valley matrix elements in each quantum dot $\Gamma_{A,B}(\theta,\varphi)$ are defined in Eq \ref{eq:SV_mtx_element}, where the phase $\eta$ may be distinct for each quantum dot. While we note that the $g$-factor may be distinct for each valley-orbital eigenstate $\vert E_{0,1}^{A,B} \rangle$, for our purposes here we neglect corrections to the $g$-factor governing the magnetic field-dependent gap between ground-ground and the ground-excited, excited-ground spin sectors.

In addition to the SOC-induced rotations of the singlet/triplet qubit, we include contributions from magnetic field gradients due to Overhauser field fluctuations $B_{\textrm{HF}}$ of contact hyperfine interaction with an evolving background nuclear spin environment, where $\delta B \to \delta B + B_{\textrm{HF}}$. Consider a spin-orbit frequency $f_{0}$ set by the applied magnetic field and $g$-factor difference. For a given shot of the experiment, fluctuations due to hyperfine interaction with nearby nuclear spins results in a random Overhauser field frequency $f$. Making the assumption that the Overhauser field is Gaussian-distributed with standard deviation $\sigma$ and is slowly varying such that a quasi-static approximation is valid, we find that the observed rotation frequency averaged over Overhauser field configurations is
\begin{eqnarray}
\langle\vert f_{0} \! + \!f\vert\rangle	& = &	\frac{1}{\sqrt{2\pi\sigma^{2}}}\int_{-\infty}^{\infty}df\ \vert f_{0} \! + \! f\vert e^{-f^{2}/2\sigma^{2}} \nonumber \\
	& = &	\sigma\sqrt{\frac{2}{\pi}}e^{-f_{0}^{2}/2\sigma^{2}} \! + \! f_{0}\mathrm{Erf}\left(\! \frac{f_{0}}{\sqrt{2\sigma^{2}}} \! \right).
    \label{eq:Overhauser_correction}
\end{eqnarray}
In the regime $\vert f_0 \vert \gg \sigma$, we see from Eq. \ref{eq:Overhauser_correction} that the Overhauser field is a small correction to the rotation frequency. For $\vert f_0 \vert \ll /\sigma$, the rotation frequency tends towards $\sigma \sqrt{2/\pi}$, being dictated fully by the Overhauser field spread.

Since applying a magnetic field can result in additional confinement normal to the applied field, for sufficiently strong magnetic fields it is possible for the valley splitting itself to depend on the magnetic field orientation. A theoretical analysis of the dependence of valley splitting with magnetic field strength is given in Ref. \cite{Friesen2006} in the context of Si/SiGe quantum dots. Since the value of the critical field can be measured precisely in our experiments and our model for SOC allows for fitting the angular dependence of the $g$-factors, we suggest that one may infer the angular dependence of the valley splittings $\Delta_{\mathrm{vs},A}(\theta,\varphi)$ and $\Delta_{\mathrm{vs},B}(\theta,\varphi)$ from this experimental protocol.

Through straightforward diagonalization of the 2x2 blocks of Eq. \ref{eq:Four-level_Hamiltonian} (with $J=0$ assumed), we find that the eigenvalues are given by $\lbrace E_{+}^{A}, E_{-}^{A}, E_{+}^{B}, E_{-}^{B} \rbrace$ with
\begin{eqnarray}
E_{\pm}^{A} & = & \frac{1}{2} \Big( \Delta_{\mathrm{vs},A} \! - \! \mu_{B} g B \pm \\
& & \sqrt{\big( \Delta_{\mathrm{vs},A} \! - \! \mu_{B} g B \big)^{2} + 4 \vert \Gamma^{A}\vert^{2} \big)} \Big) \nonumber \\
E_{\pm}^{B} & = & \frac{1}{2} \Big( \Delta_{\mathrm{vs},B} \! - \! \mu_{B} g B \pm \\
& & \sqrt{\big( \Delta_{\mathrm{vs},B} \! - \! \mu_{B} g B \big)^{2} + 4 \vert \Gamma^{B} \vert^{2} \big)} \Big) \nonumber.
\end{eqnarray}
Now, to find the rotation frequency of our $S$/$T_{0}$ qubit as a function of applied magnetic field, we need to identify the energy levels for which the two-dimensional qubit subspace $\mathrm{span}\lbrace S, T_{0} \rbrace$ has maximal support. This simply amounts to identifying, for each transition, the state that has most support on this subspace. We find that the dominant component of the rotation frequency is given by $f_{\mathrm{rot}}(\theta,\varphi) = \vert E^{A}(\theta,\varphi) - E^{B}(\theta,\varphi) \vert/h$, where
\begin{equation}
\label{eq:ENCases}
E^{N}(\theta,\varphi) = \begin{cases}
E_{-}^{N}(\theta,\varphi), & B < B_{c}^{N} \\
E_{+}^{N}(\theta,\varphi), & B > B_{c}^{N}.
\end{cases}
\end{equation}
for $B_{c}^{N} = \Delta_{\mathrm{vs,N}}/(g\mu_{B})$ the critical field for QD$_{N}$.

\subsection{Fitting to the measurements}
\label{appendix:Fits_and_statistics}
Here, we describe our procedures for analyzing the measurement data and performing the model parameter estimates. The starting dataset is the spectral power of singlet return probability as a function of manipulation time for a range of applied magnetic field strengths and orientations. Our procedure for postprocessing and fitting to the measurement data is the following:

(1) \textbf{Peak-finding}: We infer the qubit rotation frequency with a peak-finding procedure that accounts for the finite linewidth and disregards spurious noise and instrumental artifacts. We initialize our peak-finding algorithm based on a coarse manually-derived estimate of the hot spot model parameters $(\Delta_{\mathrm{vs,A}},\Delta_{\mathrm{vs,B}},\Delta g, \vert \Gamma_{A} \vert, \vert \Gamma_{B} \vert)$ for each particular magnetic field orientation.

(2) \textbf{Identification of an overall magnetic field offset}: Due to a combination of variability in the synchronization of the start of the magnetic field sweep and data collection, along with hysteresis in the magnet, for each magnetic field sweep there is a small unknown overall offset in the magnetic field strength. Since we expect the measurements to be symmetric under change of sign of the magnetic field, for the SiMOS device measurements we identify the magnetic field offset for each B-field orientation by minimizing the RMS deviation between the $f(B)$ for the positive and negative B-field intervals given an offset $B \to B + B_{\mathrm{offset}}$. The offsets determined for these experiments are given in Table. \ref{tab:BOffsets}. For our Si/SiGe dataset, we absorb this unknown magnetic field offset into our valley splitting estimates. 

\begin{figure}[!ht]
	\centering
	\includegraphics[width=0.5\textwidth]{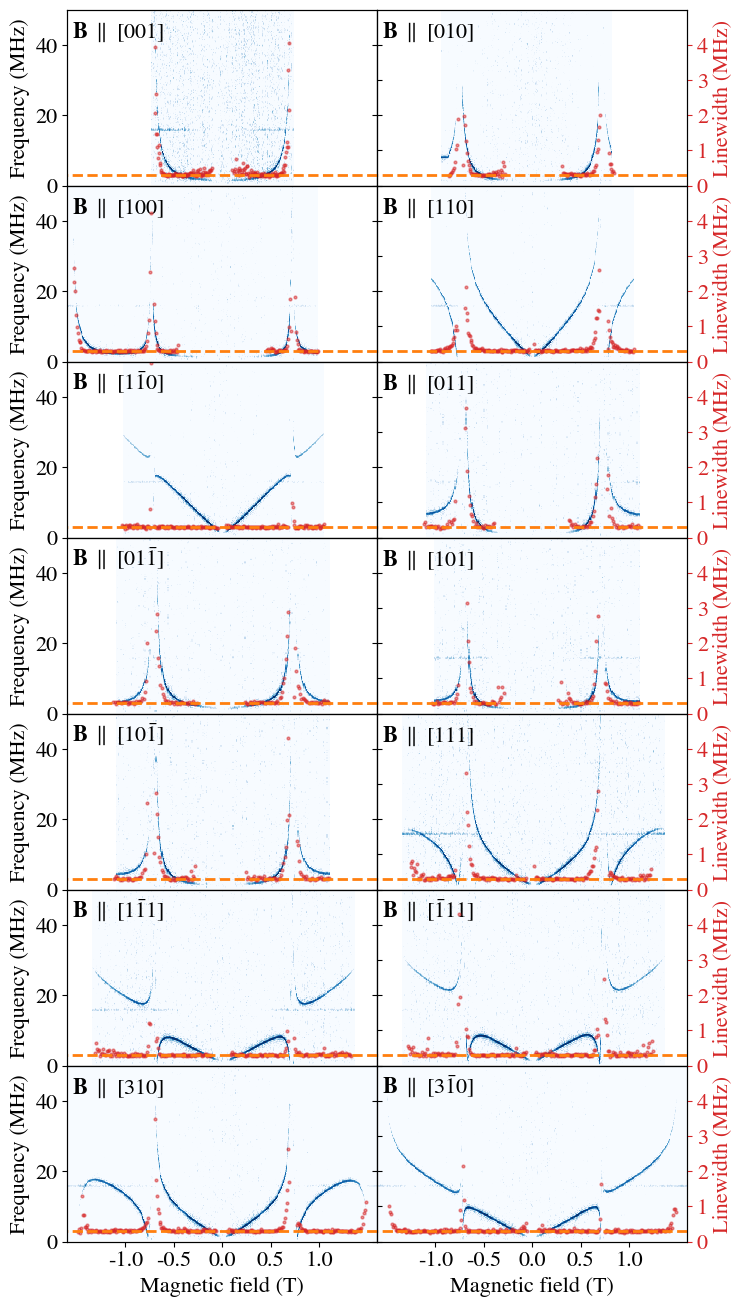}
	\caption{\textbf{Linewidths for SiMOS measurements.} Fit linewidths (red points) superimposed on the measured FFT of singlet/triplet rotation frequency for the various B-field orientations. Linewidth tends to increase in the vicinity of the hot spots, consistent with what one may expect for shorter spin relaxation times or dephasing. The orange dashed line corresponds to a baseline 300 kHz linewidth that is generally observed away from the hot spots. We suspect that shorter dephasing times near the hot spots are due to enhanced charge noise sensitivity~\cite{Jock2022}.}
	\label{fig:Linewidths_SiMOS}
\end{figure}

\begin{figure}[!ht]
	\centering
	\includegraphics[width=0.4\textwidth]{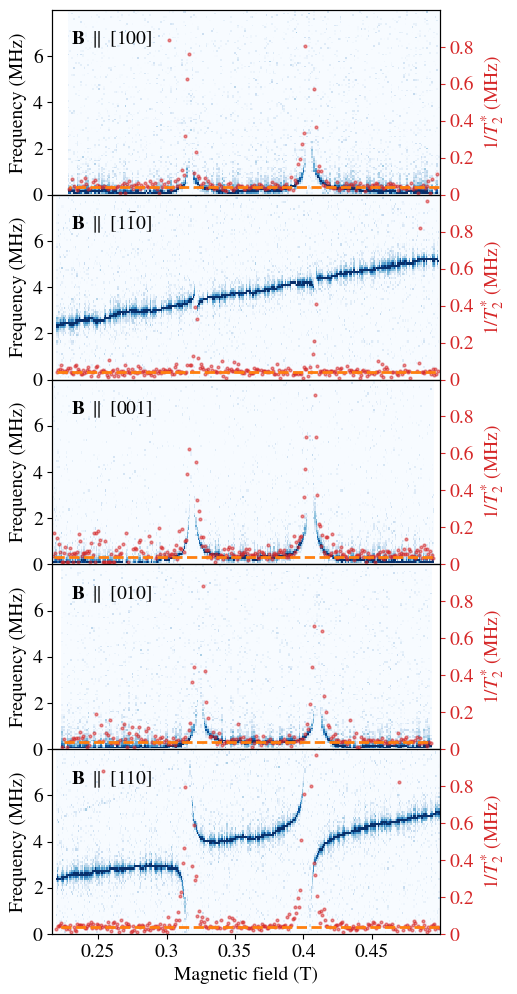}
	\caption{\textbf{Dephasing for Si/SiGe measurements.} Inverse of fit dephasing times, $1/T_{2}^{*}$ (red points) superimposed on the measured FFT of singlet/triplet rotation frequency for the various B-field orientations. As for the SiMOS measurements, coherence times appear to be shorter in the vicinity of spin-valley hot spots. The orange dashed line corresponds to a baseline 40 kHz inverse dephasing time that is generally observed away from the hot spots. We suspect that shorter dephasing times near the hot spots are due to enhanced charge noise sensitivity.}
	\label{fig:Linewidths_SiSiGe}
\end{figure}

(3) \textbf{Fitting the full model}: Next, we estimate the fixed model parameters that best agree with the full angular dependence. The input for fitting our full angular-dependent model is the set of estimated magnetic field-dependent frequency peak values $f^{\mathrm{data}}_{i}$ and associated linewidths $\sigma_{i}$ for all magnetic field orientations. For a given candidate model, the cost function we seek to minimize over the model parameters $\boldsymbol{\lambda}$ is given by $C(\boldsymbol{\lambda}) = \sum_{i} (f^{\mathrm{model}}_{i}(\mathbf{B}_{i},\boldsymbol{\lambda}) - f^{\mathrm{data}}_{i})^{2}/\sigma_{i}^{2}$, where $i$ indexes the (frequency peak, linewidth) pair at magnetic field value $\mathbf{B}_{i}$. For the Si/SiGe measurements, we minimize over the following 13 model parameters: ($\Delta_{\mathrm{vs,A}}$, $\Delta_{\mathrm{vs,B}}$, $\lbrace \Delta B_{j} \rbrace$, $\gamma_{A}$, $\gamma_{B}$, $\eta_{A}$, $\eta_{B}$, $\Delta \alpha$, $\Delta \beta$, $\sigma_{\mathrm{mag}}$), where $\lbrace B_{j} \rbrace$ are magnetic field offsets for four of the five orientations that account for valley splitting variation and/or magnet hysteresis, $\gamma_{A,B}$ are the spin-valley coupling strengths appearing in Eq. \ref{eq:SV_mtx_element}, and $\sigma_{\mathrm{mag}}$ quantifies the magnetic noise background appearing in Eq. \ref{eq:Overhauser_correction}. For the SiMOS measurements, we perform the same fitting procedure as above but over a larger number of parameters accounting for the larger number of magnetic field orientations probed. For these fits for the SiMOS dataset we include the following 22 parameters: ($\Delta_{\mathrm{vs,A}}$, $\Delta_{\mathrm{vs,B}}$, $\gamma_{A}$, $\gamma_{B}$, $\eta_{A}$, $\eta_{B}$, $\Delta \alpha$, $\Delta \beta$, $\sigma_{\mathrm{mag}}$, $\lbrace \delta \Delta_{\mathrm{vs,A}}^{j} \rbrace$), where the additional 13 parameters $\lbrace \delta \Delta_{\mathrm{vs,A}}^{j} \rbrace$ account for magnetic field-dependent shifts of the valley splitting of QD$_A$. Since the valley splitting $\delta \Delta_{\mathrm{vs,B}}$ of QD$_B$ is poorly constrained in most field orientations by the limited range of magnetic field magnitudes probed, we do not allow for it to vary with field orientation.

(4) \textbf{Uncertainty quantification}: To estimate uncertainty in our fits for all model parameters, we take the following bootstrapping approach. Given the frequency $\lbrace f_{i}^{\mathrm{data}} \rbrace$ and linewidth values $\lbrace \sigma_{i} \rbrace$ for all magnetic field values probed, we sample from a normal distribution with mean $f_{i}^{\mathrm{data}}$ and standard deviation $\sigma_{i}$ to generate a new dataset $\lbrace f_{i}' \rbrace$ for all magnetic field values $B_{i}$. We then perform a nonlinear optimization of the cost function $C'(\boldsymbol{\lambda}) = \sum_{i} (f^{\mathrm{model}}_{i}(\mathbf{B}_{i},\overrightarrow{\lambda}) - f_{i}')^{2}$ to generate a point estimate $\boldsymbol{\lambda}'$ of the model parameters. We then repeat this 200 times. Our reported uncertainty for each parameter in Table \ref{tab:results_table} is the standard deviation of the corresponding best-fit values.

\section{Additional measurements}
\label{appendix:Additional_measurements}
\begin{figure*}[t]
	\centering
    \includegraphics[width=0.75\textwidth]{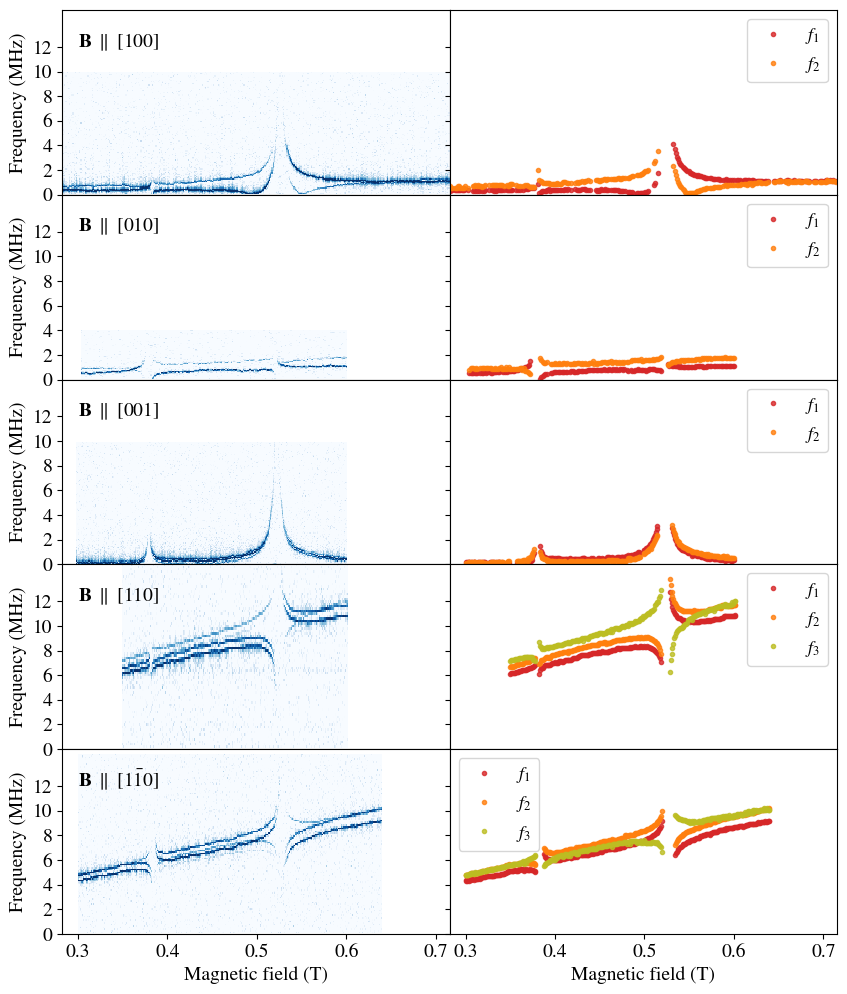}
	\caption{\textbf{Measurements in a a second pair of QDs, (QD$_2$,QD$_3$), of the Si/SiGe triple QD.} (left column) Magnetic field sweeps of ST free induction decay measurements for the same set of five magnetic field orientations in which multiple frequency components are evident. (right column) Plots of the corresponding extracted peak frequencies, with frequency component inferred by tracking the prominence of the FFT signal in the left column data. Note that in only the last two field orientations are three frequencies evident.}
    \label{fig:ExperimentalMagnetospectroscopy_SiSiGe2}
\end{figure*}

In Fig. \ref{fig:ExperimentalMagnetospectroscopy_SiSiGe2}, we show the results of similar measurements for a second pair of QDs in the same Si/SiGe triple QD device. These measurements were carried out in the same manner as for the data featured in Fig.~\ref{fig:ExperimentalMagnetospectroscopy_SiSiGe}, but the double QDs were now formed underneath the center and rightmost plunger gates with respect to Fig.~\ref{fig:DeviceIntro}c. This was accomplished by flooding the leftmost plunger gate to extend the electron reservoir out to the second barrier gate from the left. A notable difference of these measurements compared to those discussed in the main text is the appearance of multiple frequency components. As many as four frequencies may be present in the $[1\bar{1}0]$ orientation, though we could reliably perform peak finding for only two or three frequency components per magnetic field scan. 

From individual fits to the hot spot B-field magnitude dependence, we extracted the spin-valley coupling for the dot A and B transitions as well as the corresponding valley splittings. Noting that the voltage tuning conditions for these measurements of the other QD pair are different from the results of the main text, we can see that both the spin-valley coupling and valley splitting for QD$_{B}$ is significantly larger than for QD$_{A}$, with the valley splitting of QD$_{A}$ more consistent with QD$_A$ and QD$_{B}$ for the measurements of the QD pair in the main text. This may suggest that the QD under the rightmost plunger gate may have the higher valley splitting. We suggest that future work exploring spin-valley hot spots with intentionally prepared excited valley eigenstate population may reveal further insights into the interplay of valley physics and SOC at interfaces. This can be done, for example, by varying the singlet state preparation dynamics. We show this in Fig.~\ref{fig:ExcitedValleys}, where the ramp time from $S(4,0)$ to $S(3,1)$ is varied while the wait time is ramped in a free induction decay experiment. The resulting FFT shows three ST rotation frequencies that vanish in decreasing energy order as the ramp time increases. This is a simple demonstration of controllably populating excited valley states by manipulating the adiabaticity of singlet state preparation that can be used to expand the study of valley physics.

\begin{table}[ht]
\begin{tabular}{| c | c |}
\hline
\textbf{Parameter} & \textbf{Si/SiGe (QD$_2$,QD$_3$)} \\
\hline
Valley splitting, $\Delta_{\mathrm{vs,A}}$ ($\mathrm{\mu eV}$)  & 44.0(2)  \\
\hline
Valley splitting, $\Delta_{\mathrm{vs,B}}$ ($\mathrm{\mu eV}$) & 60.7(1) \\
\hline
Spin-valley coupling, $\gamma_{\mathrm{A}}$ ($\mathrm{\mu eV}$) & 0.045 \\
\hline
Spin-valley coupling, $\gamma_{\mathrm{B}}$ ($\mathrm{\mu eV}$) & 0.11 \\
\hline
\end{tabular}
\caption{\textbf{Fit valley splitting and SOC parameters for measurement of separate dot pair of Si/SiGe device exhibiting multiple frequencies.} Our reported error bars for the valley splitting are the standard deviation of the point estimates of valley splittings from the five measured B-field orientations. We report the point estimate for the spin-valley coupling based on the orientation, [001], that exhibits maximal spin-valley coupling.}
\label{tab:multi_frequencyresults_table}
\end{table}

\appendix
\bibliography{bibliography}
\bibliographystyle{unsrt}
\end{document}